\documentclass[aps,pre,twocolumn, epsfig]{revtex4}
\usepackage{amssymb}
\usepackage{amsbsy}
\usepackage{amsmath}
\usepackage{epsfig}

\begin{document}

\title{Exact analytical evaluation of time dependent transmission coefficient from
the method of reactive flux for an inverted parabolic barrier}
\author{Rajarshi Chakrabarti}
\address{Department of Inorganic and Physical Chemistry, Indian
Institute of Science, Bangalore, 560012, India}

\begin{abstract}
In this paper we derive a general expression for the transmission
coefficient using the method of reactive flux for a particle coupled
to a harmonic bath surmounting a one dimensional inverted parabolic
barrier. Unlike Kohen and Tannor [J. Chem. Phys. \textbf{103}, 6013
(1995)] we use a normal mode analysis where the unstable and the
other modes have a complete physical meaning. Importantly our
approach results a very general expression for the time dependent
transmission coefficient not restricted to overdamped limit. Once
the spectral density for the problem is known one can use our
formula to evaluate the time dependent transmission coefficient. We
have done the calculations with time dependent friction used by Xie
[Phys. Rev. Lett \textbf{93}, 180603 (2004)] and also the one used
by Kohen and Tannor [J. Chem. Phys. \textbf{103}, 6013 (1995)]. Like
the formula of Kohen and Tannor our formula also reproduces the
results of transition state theory as well as the Kramers theory in
the limits $t\rightarrow 0$ and $t\rightarrow \infty$ respectively.
\end{abstract}

\maketitle

\address{Department of Inorganic and Physical Chemistry, Indian
Institute of Science, Bangalore, 560012, India}

\affiliation{Department of Inorganic and Physical chemistry, Indian
Institute of Science, Bangalore 560012, India}

\section{Introduction}

The dynamics of a particle surmounting a barrier is an important and
interesting problem in chemistry and physics. In chemistry, chemical
reactions are most common examples where one studies the dynamics of barrier
crossing along a suitably defined reaction coordinate. While in physics
nucleation phenomena, electrical transport etc. involve barrier crossing. A
review by H\"anggi, Talkner and Borkovec \cite{HanggiRMP1990} gives a
comprehensive account on the subject of barrier crossing. The simplest
possible approach to calculate the rate constant for such processes is given
by transition state theory \cite{HanggiRMP1990, NitzanBook}. As the simplest
version of the problem one can describe the rate process along a single
reaction co-ordinate. Then the one dimensional version of the transition
state rate constant $k_{TST}$ reads as, $k_{TST}=\frac{\omega _0}{2\pi }\exp
[-\frac{E_{b}}{k_BT}]$, where $\omega_0$ is the angular frequency in the
reactant well and $E_{b}$ is the barrier height or the activation energy.
The derivation of transition state rate constant is based on two
assumptions. Firstly an overall thermal equilibrium is assumed and secondly
frictional effect is not taken into account by incorporating what is known
as ``no recrossing" assumption. Subsequently, Kramers \cite{HanggiRMP1990,
KramersPhysica1940, CoffeyBook, NitzanBook} theory takes into account of the
friction in a phenomenological way. His formulation was based on Markovian
dynamics of a Brownian particle escaping from a metastable state. Further
extensions of Kramers's theory were carried out by Grote and Hynes \cite
{GroteJCP1980} followed by H\"anggi and Mojtabi \cite{MojtabiPRA1986}. They
took into account of the non-Markovian nature of the dynamics. Another
approach to calculate the rate constant is due to Chandler \cite
{ChandlerJSP1986, ChandlerBook} known as the method of reactive flux in
chemical physics literature. In this approach the thermal rate constant is
written as a correlation function which is nothing but the ensemble average
over infinite number of trajectories starting at the barrier top and ending
on the product side at time $t$. The rate constant at time $t$ is written as
a product of the time dependent transmission coefficient $\kappa(t)$ and the
transition state rate constant $k_{TST}$. Thus one writes, $k(t)=\kappa(t)
k_{TST}$. In the limit $t\rightarrow 0$, $\kappa(t)\rightarrow1$ and one
obtains $k(0)=k_{TST}$. Here we follow the method of reactive flux to
calculate the time dependent rate constant for a particle bilinearly coupled
to a harmonic bath \cite{WeissBook, ZwanzigBook, ZwanzigJPCA}. The bilinear
coupling allows us to get an analytical expression for the time dependent
transmission coefficient. Although the method of reactive flux and the
hamiltonian we have used are well known in the literature, as far as our
knowledge nobody has used such hamiltonian to calculate the time dependent
transmission coefficient analytically. The calculation is of interest as the
transmission coefficient for the overdamped limit of the model was obtained
recently by simulation \cite{XiePRE2006}. Our approach directly gives
expression for the time dependent transmission coefficient in terms of the
spectral density $J(\omega)$ and not restricted to overdamped limit. A
normal mode analysis of the coupled Hamiltonian was made by Pollak \cite
{PollakJCP1986, PollakJCP1989, PollakJCP1988} and others to analyze the
Kramers problem. Here we point out that the approximation is very powerful
and can be used to get the exact expression for $\kappa(t)$. Further we
believe that it can also be used to get the quantum expression for $%
\kappa(t) $. An exact calculation of $\kappa(t)$ using the phase space
distribution function formulation was carried out in classical regime by
Kohen and Tannor \cite{TannorJCP1995, TannorACP2000}, Bao \cite{BaoJCP2006},
and more recently in the context of single enzyme kinetics by Chowdhury and
Cherayil \cite{CherayilJCP2006}. Also the calculation of the transmission
coefficient in the quantum domain is carried out by Ray's group \cite
{DSRAYJCP2003}. Our analysis in terms of normal modes has the advantage that
it makes the physical ideas very clear. The paper is arranged as follows. In
section II we introduce our system plus harmonic bath hamiltonian and the
Generalized Langevin Equation (GLE). In section III we mention the method of
reactive flux and using our hamiltonian derive an analytical expression for
the time dependent transmission coefficient, $\kappa(t)$. Section IV deals
with the calculation of $\kappa(t)$ for Markovian and Non Markovian
dynamics. Section V is conclusion.

\section{Generalized Langevin Equation and normal mode analysis}

We consider a particle coupled to a harmonic bath \cite{WeissBook,
ZwanzigBook} with the total hamiltonian

\begin{equation}
\label{hamiltonian1}H=\frac{P^2}{2M}+V(Q)+\frac 12\sum\limits_{j=1}^N\left(
\frac{P_j^2}{m_j}+m_j\omega _j^2\left( Q_j-\frac{c_j}{m_j\omega _j}Q\right)
^2\right)
\end{equation}

Here ($P_j, Q_j$) are the momenta and coordinates of the $j$th bath
oscillator whose mass and frequency are $m_j, \omega_j$ respectively. ($P, Q$%
) is the momentum and coordinate of the system. $c_j$ couples the bath
oscillator to the system. Now in terms of the mass weighted coordinates

\begin{equation}
\label{transformation}q=\sqrt{M}Q,q_j=\sqrt{m_j}Q_j
\end{equation}

our Hamiltonian becomes

\begin{equation}
\label{hamiltonian2}H=\frac{p^2}2+V(q)+\frac 12\sum\limits_{j=1}^N\left(
p_j^2+\omega _j^2\left( q_j-\frac{c_j}{\sqrt{Mm_j}}q\right) ^2\right)
\end{equation}
and then with this Hamiltonian one can derive what is know as the
Generalized Langevin Equation (GLE) which reads as

\begin{equation}
\label{GLE}\ddot q(t)+V^{^{\prime }}(q)+\int\limits_0^tdt^{^{\prime }}\gamma
(t-t^{^{\prime }})\dot q(t^{^{\prime }})=\zeta (t)
\end{equation}

One can also write the time dependent friction in terms of the spectral
density of the bath $J(\omega)$ as follows

\begin{equation}
\label{frictionpower}\gamma (t)=\frac{\Theta (t)}{M}\frac 2\pi
\sum\limits_{j=1}^N\int\limits_0^\infty d\omega \frac{J(\omega )}\omega \cos
(\omega t)
\end{equation}

where

\begin{equation}
\label{power}J(\omega )=\frac \pi 2\sum\limits_{j=1}^N\frac{c_j^2}{m_j\omega
_j}\delta (\omega -\omega _j)
\end{equation}

One can easily show that

\begin{equation}
\label{noisecorr1}\left\langle \zeta (t_1)\zeta (t_2)\right\rangle =M
k_BT\gamma (t_1-t_2)
\end{equation}
We take the barrier top to be at $q=0$ and introduce $\omega _b^2=\left(
\frac{\partial ^2V}{\partial q^2}\right) _{q=0}$. For small amplitude motion
around the barrier top, one can use the normal modes, $\eta
_j=\sum\limits_{i=0}^NU_{ji}q_i $ where $\underline{\underline U}$ is an
orthogonal matrix such that $\underline{\underline U}$ $\underline{%
\underline D}$ $\underline{\underline U}^T$ is diagonal. $\underline{%
\underline D}$ is a dynamical matrix given by

\begin{equation}
\label{heissian}\underline{\underline D}=\left(
\begin{array}{cccc}
-\omega _b^2+\sum\limits_{j=1}^N\frac{c_j^2}{Mm_j\omega _j^2} & . & . & -
\frac{c_j}{\sqrt{Mm_j}} \\ . & . & . & 0 \\
. & . & . & 0 \\
-\frac{c_j}{\sqrt{Mm_j}} & 0 & 0 & \omega _j^2
\end{array}
\right)
\end{equation}

The reaction co-ordinate $q$ may then be written as
$$
q=\sum\limits_{j=0}^NU_{0j}\eta_i
$$

Of the modes $\eta_j$, we take $\eta_0$ to be the unstable mode. We denote
the frequency of $\eta_j$ as $\lambda_j$. $\eta_0$ has the imaginary
frequency and we write it as $\lambda_0=i \Lambda$, where $\Lambda$ is real.
As $\eta_j$ is a normal mode, its time development is given by
$$
\eta_j(t)=\eta_0(0)\cos(\lambda_j t)+ p_{\eta_j}(0)\frac {\sin (\lambda_j
t)}{\lambda_j}%
$$
with $j=0,1,2,.....N$.

Now the Hamiltonian when written in terms of the normal modes the coupling
is no more there and is given by

\begin{equation}
\label{hamiltoniannormal}H_{nor}=\sum\limits_{j=0}^N\left( \frac{p_j^2}%
2+\frac 12\lambda _j^2\eta _j^2\right)
\end{equation}

In the next section we analytically derive the rate constant in the normal
mode description using the above Hamiltonian.

\section{Method of reactive flux and the rate constant}

In this section we briefly describe the method of reactive flux \cite
{ChandlerJSP1986, ChandlerBook} to calculate the rate constant. The method
of reactive flux expresses the thermal rate constant (written in the
mass-weighted coordinates) as

\begin{equation}
\label{rateconstant2}k(t)=\frac{\left\langle \dot q(t)\delta (q(0))\theta
_P(q(t))\right\rangle }{\left\langle \theta _R(q(0))\right\rangle }=\frac
{k_n(t)}{k_d(t)}
\end{equation}

where the angular bracket indicates the thermal average over the system as
well as the bath degrees of freedom. The top of the barrier is chosen as $%
q(0)=0$. $\theta_R(q(0))$ is 1 if $q(0)<1$ and $0$ otherwise. $\theta_p$ is
just 1-$\theta_R$, $\dot q$ is the velocity of the particle.

The numerator of Eq. (\ref{rateconstant2}) could be written as

\begin{equation}
\label{numera1}k_n(t)=-i\left( \frac \partial {\partial r_0}\left\langle
\exp (ir_0\dot q(0))\delta (q(0))\theta _P(q(t))\right\rangle \right)
_{r_0=0}
\end{equation}

Expressing the step function as an integral over delta function and using
the Fourier integral representation of delta function the expression for $%
k_n(t)$ becomes

\begin{widetext}
\begin{eqnarray*}
k_n(t)=\left( \frac 1{\left( 2\pi \right) ^2}\int\limits_0^\infty
dh(-i)\frac \partial {\partial r_0}\int\limits_{-\infty }^\infty
ds_0\int\limits_{-\infty }^\infty ds_1\exp (-is_1h)\left\langle \exp
(ir_0\dot q(0)+is_1q(t)+is_0q(0))\right\rangle \right) _{r_0=0}
\end{eqnarray*}
\end{widetext}

The angular bracket indicates the thermal average over all degrees of
freedom of the problem and it is done by multiplying the quantities within
the angular bracket by $\exp (-\beta H_{nor})$ , and carrying out the
integration over all coordinates and momenta degrees of freedom. Writing $%
k_n(t)$ in terms of $\eta_j$ leads to

\begin{widetext}
\begin{equation}
\label{kndetails}k_n(t)=\frac{\exp (-\beta E_b)}{\left( 2\pi \right) ^2}%
\left(
\begin{array}{c}
\int\limits_0^\infty dh(-i)\frac \partial {\partial
r_0}\int\limits_{-\infty }^\infty ds_0\int\limits_{-\infty }^\infty
ds_1\exp (-is_1h)\exp (-\beta \left( \frac{p_{\eta
_0}^2}2-\frac{\Lambda ^2}2\eta _0^2\right) )\exp (ir_0U_{00}p_{\eta
_0}) \\ \exp (ir_0U_{00}\eta _0)\exp (is_1\left( U_{00}\eta _0\cosh
(\Lambda t)+p_{\eta _0}(0)
\frac{\sinh (\Lambda t)}\Lambda \right) ) \\ \prod\limits_{j=1}^N\int%
\limits_{-\infty }^\infty dp_j\int\limits_{-\infty }^\infty d\eta
_j\exp (-\beta \left( \frac{p_j^2}2+\frac{\lambda _j^2}2\eta
_j^2\right) )\exp (ir_0U_{j0}\eta _j)\exp (ir_0U_{j0}p_{\eta _j}) \\
\exp (is_1\left( U_{j0}\eta _j\cos (\lambda _jt)+p_{\eta
_j}(0)\frac{\sin (\lambda _jt)}{\lambda _j}\right)
\end{array}
\right) _{r_0=0}
\end{equation}
\end{widetext}
where $E_b$ is the barrier height. After one carries out the integrations
the expression for $k_n(t)$ can be written as
\begin{widetext}
\begin{equation}
\label{knfinal}k_n(t)=\prod\limits_{j=1}^N\left( \frac{2\pi }{\beta
\lambda _j}\right) \left( \frac 1{\beta \Lambda }\right)
\left( \frac{-\dot c(t)}{\left( c(t)^2-c(0)^2\right) ^{\frac 12}}\right) \exp (-\beta E_b)
\end{equation}
\end{widetext}

Similarly the denominator of the Eq. (\ref{rateconstant2}) can be written as
\begin{widetext}
\begin{equation}  \label{kdfinal}
k_d(t)=\prod\limits_{j=0}^N\int\limits_{-\infty }^\infty
dp_j\int\limits_{-\infty }^\infty d\eta _j\exp (-\beta \left( \frac{p_j^2}2+%
\frac{\lambda _j^{0^2}}2\eta _j^2\right) )=\prod\limits_{j=0}^N\left( \frac{2\pi }{\beta \lambda _j^0}\right)
\end{equation}
\end{widetext}

Then the time dependent rate constant becomes

\begin{equation}
\label{kfinal}k(t)=\frac{k_n(t)}{k_d(t)}=\frac 1{2\pi \Lambda }\frac{%
\prod\limits_{j=0}^N\lambda _j^0}{\prod\limits_{j=1}^N\lambda _j}\left(
\frac{-\dot c(t)}{\left( c(t)^2-c(0)^2\right) ^{\frac 12}}\right) \exp
(-\beta E_b)
\end{equation}

Here we use the following identity \cite{HanggiRMP1990}, $\frac{%
\prod\limits_{j=0}^N\lambda _j^0}{\Lambda \prod\limits_{j=1}^N\lambda _j}=%
\frac{\omega _0}{\omega _b}$, and this simplifies our $k(t)$

\begin{equation}
\label{ktsimple}k(t)=\frac 1{2\pi }\frac{\omega _0}{\omega _b}\left( \frac{%
-\dot c(t)}{\left( c(t)^2-c(0)^2\right) ^{\frac 12}}\right) \exp (-\beta
E_b)
\end{equation}
where, $c(t)=\left( \sum\limits_{j=0}^N\frac{U_{0j}^2\cos
(\lambda _jt)}{\lambda _j^2}%
\right)$. As usual $k(t)$ \cite{TannorJCP1995} is defined by

\begin{equation}
\label{ktkappat}k(t)=\kappa (t)k(0)
\end{equation}

We show in the Appendix that $k(0)=\frac{\omega _0}{2\pi }\exp (-\beta E_b)$%
, which is nothing but the transition state rate constant $k_{TST}$. In the
Appendix we also show how one gets Kramers rate constant in the $%
t\rightarrow \infty$ limit.

Using the above expression for $k(0)$ our time dependent transmission
coefficient, $\kappa (t)$ becomes
\begin{equation}
\label{kappat}\kappa (t)=\frac 1{\omega _b}\left( \frac{-\dot c(t)}{\left(
c(t)^2-c(0)^2\right) ^{\frac 12}}\right)
\end{equation}

In the next section we calculate $c(t)$ and from that $\kappa(t)$ for
Markovian as well as for Non Markovian noise.

\section{Calculation of $c(t)$ and $\kappa (t)$}

Calculation of $\kappa(t)$ involves the calculation of $c(t)$. First we
describe a general derivation of $c(t)$ then we evaluate $\kappa(t)$ for
different cases.

\subsection{Calculation of $c(t)$}

$c(t)$ is defined by $c(t)=\sum\limits_{j=0}^N\frac{U_{j0}^2\cos
(\lambda _jt)}{\lambda _j^2}$. One proceeds by writing
$\sum\limits_{j=1}^NU_{j0}^2\delta (\omega ^2-\lambda _j^2)=-\frac
1\pi \text{Im}\sum\limits_{j=1}^N\frac{U_{j0}^2}{\left(
\omega ^2+i\eta -\lambda _j^2\right) }.$ To evaluate $\sum\limits_{j=1}^N%
\frac{U_{j0}^2}{\left( \omega ^2+i\eta -\lambda _j^2\right) }$ here
we adopt notations similar to quantum mechanics and also use the
fact that $D$ is a matrix with eigen values $\lambda
_j^2,j=1,2,.....N.$

\begin{widetext}
\begin{equation}
\label{sumqm}\sum\limits_{j=1}^N\frac{U_{j0}^2}{\left( \omega
^2+i\eta -\lambda _j^2\right) }=\sum\limits_{j=1}^N\left\langle
0\mid j\right\rangle \frac 1{\left( \omega ^2+i\eta -\lambda
_j^2\right) }\left\langle j\mid 0\right\rangle
=\sum\limits_{j=1}^N\left\langle 0\mid j\right\rangle \frac 1{\left(
\omega ^2+i\eta -D\right) }\left\langle j\mid 0\right\rangle
\end{equation}
\end{widetext}

Then by introducing $\left| j\right\rangle \left\langle j\right| $
in
between the sum and using the resolution of identity $\sum\limits_{j=1}^N$ $%
\left| j\right\rangle \left\langle j\right| =I$.

\begin{widetext}
\begin{equation}
\label{rowomega}\sum\limits_{j=1}^NU_{j0}^2\delta (\omega ^2-\lambda
_j^2)=\left\langle 0\right| \frac 1{\left( \omega ^2+i\eta -D\right)
}\left| 0\right\rangle =-\frac 1\pi \text{Im} \left(
\underline{\underline{G}}(\omega ^2+ i \eta)\right)
\end{equation}
\end{widetext}

\begin{figure}[tbp]
\centering \epsfig{file=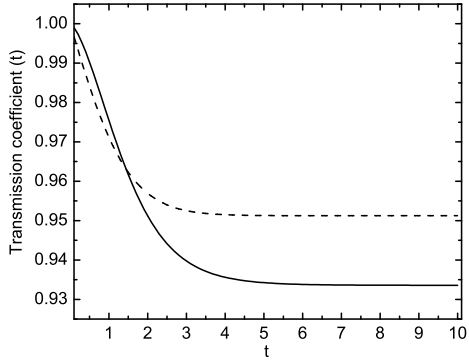,width=.8\linewidth}\newline .
\caption{Time dependent transmission coefficient,
$\protect\kappa(t)$ against time, for $H=\frac 1 2$ (dashed line)
and $H=\frac 3 4$ (solid line), parameters used are
$\protect\omega_b=1, \protect\gamma=0.1$} \label{kappa1}
\end{figure}

\begin{figure}[tbp]
\centering \epsfig{file=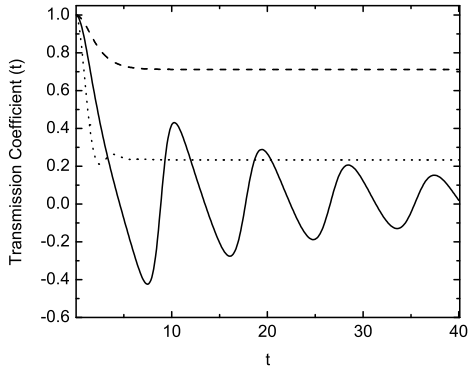,width=.8\linewidth}\newline
\caption{Time dependent transmission coefficient,
$\protect\kappa(t)$
against time for Non-Markovian friction, $\protect\gamma (t)=\protect\gamma h\exp \left( -ht\right) $. Three different regimes, caging regime (solid
line) with parameters, $\protect\gamma=150, h=0.01, \protect\omega_b=1$,
intermediate regime (dotted line) with parameters $\protect\gamma=5, h=1,
\protect\omega_b=1$, and nonadiabatic regime (dashed line) with parameters $\protect\gamma=50, h=0.01, \protect\omega_b=1$. }
\label{kappakohen}
\end{figure}

Using partitioning technique one can evaluate $(\underline{\underline{G}}%
)_{00}$ to get
\begin{equation}
\label{g002}
\begin{array}{c}
\left(
\underline{\underline{G}}(\omega ^2+i\eta )\right) _{00} \\ =\left( \omega
^2+\omega _b^2-\frac 1M\sum\limits_{j=1}^N\frac{c_j^2\omega ^2}{m_j\omega
_j^2\left( \omega ^2+i\eta -\omega _j^2\right) }\right) ^{-1}
\end{array}
\end{equation}

Then using the definition of $J(\omega )$ Eq.(\ref{power}) it becomes
\begin{equation}
\label{g00j}
\begin{array}{c}
\left(
\underline{\underline{G}}(\omega ^2+i\eta )\right) _{00} \\ =\left( \omega
^2+\omega _b^2-\frac{2\omega ^2}{M\pi }\int\limits_0^\infty \frac{J(\tilde
\omega )d\tilde \omega }{\tilde \omega \left( \omega ^2+i\eta -\tilde \omega
^2\right) }\right) ^{-1}
\end{array}
\end{equation}

As $-\frac 1\pi \text{Im} \left( \underline{\underline{G}}(\omega
^2)\right) _{00} $ $=\sum\limits_{j=0}^NU_{j0}^2\delta (\omega
^2-\lambda _j^2)=\rho (\omega ^2).$ We can write
$c(t)=2\int\limits_0^\infty \frac{d\omega }\omega \cos (\omega
t)\rho (\omega ^2).$

\subsection{Calculation of $\kappa (t) $ with $\gamma (t)=\gamma 2H\left(
2H-1\right) \left| t\right| ^{2H-2}$}

First we perform the calculation with the time dependent friction $\gamma(t)$
used by Xie \cite{XiePRL2004}, which is

\begin{equation}
\label{friction}\gamma (t)=\gamma 2H\left( 2H-1\right) \left| t\right|
^{2H-2}
\end{equation}
with, $\frac 12<H<1 $, thus the thermal fluctuations are described by
fractional Gaussian noise \cite{XiePRL2004}. Now we take a inverse Fourier
transform of Eq. (\ref{frictionpower}) and write $J(\omega)$ in terms of $%
\gamma(t)$.

\begin{equation}
\label{jomega}J(\omega )=M \omega \int\limits_0^\infty dt\gamma (t)\cos
(\omega t)
\end{equation}

Using the definition of $\gamma(t)$ we get

\begin{equation*}
\frac {2 \omega ^2}{M \pi} \int\limits_0^\infty \frac{J(\tilde \omega
)d\tilde \omega }{\tilde \omega \left( \omega ^2+i\eta -\tilde \omega
^2\right) }=\gamma \Gamma (2H+1)\Lambda ^{2-2H}
\end{equation*}

First we consider the case $\omega^2=-\Lambda^2$ which means we consider the
unstable mode of the problem. Then

\begin{equation}
\label{g00lamda}\left( \underline{\underline{G}}(-\Lambda ^2)\right)
_{00}=\left( -\Lambda ^2+\omega _b^2-\gamma \Gamma (2H+1)\Lambda
^{2-2H}\right) ^{-1}
\end{equation}

Our next task is to find out the poles, i.e. the unstable modes for which $%
G_{00}\left( -\Lambda ^2\right)$ blows up, i.e. the solutions of the
equation
\begin{equation}
\label{eqlambda}-\Lambda ^2+\omega _b^2-\gamma \Gamma (2H+1)\Lambda
^{2-2H}=0
\end{equation}
On the other hand, to find $\rho (\omega ^2) $ we have to consider a
situation where $\omega^2$ is positive. In other words the stable modes. Now
the Evaluation of $\rho (\omega ^2) $ involves the integral $\frac{ 2\omega
^2}{M \pi} \int\limits_0^\infty \frac{J(\tilde \omega )d\tilde \omega }{%
\tilde \omega \left( \omega ^2+i\eta -\tilde \omega ^2\right) } $.
Using the identity $\frac 1{\left( \omega ^2+i\eta -\tilde \omega
^2\right) }=P\left( \frac 1{\omega ^2-\tilde \omega ^2}\right) -i\pi
\delta \left( \omega ^2-\tilde \omega ^2\right)$, where $P\left(
\frac 1{\omega ^2-\tilde \omega ^2}\right) $ is the principal value
of $\left( \frac 1{\omega ^2-\tilde \omega ^2}\right) $ we find

\begin{equation}
\label{integral}\frac{2\omega ^2}{M\pi }\int\limits_0^\infty \frac{J(\tilde
\omega )d\tilde \omega }{\tilde \omega \left( \omega ^2+i\eta -\tilde \omega
^2\right) }=-\exp \left( iH\pi \right) \Gamma (2H+2)\omega ^{2-2H}\gamma
\end{equation}
Thus
\begin{equation}
\label{g00positive}\left( \underline{\underline{G}}(\omega ^2)\right)
_{00}=\left( \omega ^2+\omega _b^2+\exp \left( iH\pi \right) \Gamma
(2H+2)\omega ^{2-2H}\gamma \right) ^{-1}
\end{equation}
and

\begin{widetext}
\begin{equation}
\label{rhoomega}
\begin{array}{c}
\rho (\omega ^2)=-\frac 1\pi
\text{Im}\left( \underline{\underline{G}}(\omega ^2)\right) _{00} \\  \\
=\frac 1\pi \frac{\gamma \Gamma (2H+1)\omega ^{2-2H}\sin (H\pi )}{\left(
\left( \omega ^2+\omega _b^2+\gamma \Gamma (2H+1)\omega ^{2-2H}\cos (H\pi
)\right) ^2+\gamma ^2\sin ^2(H\pi )\Gamma (2H+1)^2\omega ^{4-4H}\right) }
\end{array}
\end{equation}
\end{widetext}

\subsubsection{The Case $H=\frac 12$}

First we consider $H=\frac 12$, for which the $\gamma(t)=2\gamma \delta(t)$
and with this the noise-noise correlation function defined in Eq. (\ref
{noisecorr1}) becomes delta function correlated, reading
\begin{equation}
\label{deltacorr}\left\langle \zeta (t_1)\zeta (t_2)\right\rangle =2M\gamma
k_BT\delta (t_1-t_2)
\end{equation}
Then the dynamics is described by usual Langevin equation rather than a GLE.
Thus the dynamics is then Markovian i.e. with no memory. For the Markovian
case Eq. (\ref{eqlambda}) becomes, $-\Lambda ^2+\omega _b^2-\gamma \Lambda
=0 $ and has two roots. But only one of them is physically acceptable. The
root is, $\Lambda _{H=\frac 12}=-\frac \gamma 2+\sqrt{\left( \frac \gamma
2\right) ^2+\omega _b^2} $. Similarly from Eq. (\ref{rhoomega}) one gets $%
\rho (\omega ^2)_{H=\frac 12}=\frac{\omega \gamma }{\pi \left( \omega
^2\gamma ^2+\left( \omega ^2+\omega _b^2\right) ^2\right) }$. Then $\Lambda
_{H=\frac 12}$ and $\rho (\omega ^2)_{H=\frac 12}$ are used to calculate $%
c(t)$ and taking $t\rightarrow 0$ limit one gets $c(0)$. In this case we
could do these analytically. The time dependent transmission coefficient in
this case is given by
\begin{widetext}
\begin{equation*} \kappa (t)=\frac{\left(
-1+e^{t(f-\gamma )}\right) }{\sqrt{\left( 1-4e^{ft}+2e^{t(f-\gamma
)}+2e^{2t(f-\gamma )}\right) -(\frac \gamma {2\omega
_b})(f-fe^{2t(f-\gamma )}+2\gamma (-1+e^{ft}))}}
\end{equation*}
\end{widetext}
where $f=\gamma+ {(4 {\omega_b}^2+\gamma)}^{1/2}$.

$\kappa(t)$ is then calculated using the above expression and plotted
against time (Fig. 1).

\subsubsection{The Case $H=\frac 34$}

When $H\neq \frac 12 $ then the noise is no longer delta function
correlated, i.e. not a white noise with Gaussian distribution. This regime
is know as Non Markovian regime which has memory. We have similarly
evaluated $\kappa(t)$ for $H =\frac 34 $ and plotted against time in Fig. 2.
But in this case the integrations are done numerically since analytical
evaluation of integrals were not possible.

\subsection{Calculation of $\kappa (t) $ with $\gamma (t)=\gamma h\exp
\left( -ht\right) $}

In this section we carry out the calculation of $\kappa(t)$ with a time
dependent friction $\gamma (t)=\gamma h\exp \left( -ht\right)$. The same
friction is used by Kohen and Tannor \cite{TannorJCP1995, TannorACP2000}.
First we calculate $J(\omega)$ with this friction as follows
\begin{equation*}
J(\omega )=M \gamma h\omega \int\limits_0^\infty \exp \left( -ht\right) \cos
\left( \omega t\right) dt=\frac{\gamma Mh^2\omega }{\left( h^2+\omega
^2\right) }
\end{equation*}

Then the following integral is evaluated as follows
\begin{widetext}
\begin{equation*} \frac {2\omega ^2}{M \pi}
\int\limits_0^\infty \frac{J(\tilde \omega )d\tilde \omega }{\left(
\alpha ^2+\tilde \omega ^2\right) \left( \omega ^2+i\eta -\tilde
\omega ^2\right) }=\frac{2\omega ^2h^2}\pi \int\limits_0^\infty
\frac{d\tilde \omega }{\left( \alpha ^2+\tilde \omega ^2\right)
\left( \omega ^2+i\eta -\tilde \omega ^2\right) }=\frac{\gamma
h\omega \left( \omega -ih\right) }{\left( h^2+\omega ^2\right) }
\end{equation*}
\end{widetext}

We have used the identity, $\frac 1{\left( \omega ^2+i\eta -\tilde
\omega ^2\right) }=P\left( \frac 1{\omega ^2-\tilde \omega
^2}\right) -i\pi \delta \left( \omega ^2-\tilde \omega ^2\right)$,
where $P\left( \frac 1{\omega ^2-\tilde \omega ^2}\right) $ is the
principal value of $\left( \frac 1{\omega ^2-\tilde \omega
^2}\right) $. Thus

$$
\left( \underline{\underline{G}}(\omega ^2)\right) _{00}=\left( \omega
^2+\omega _b^2-\frac{\gamma h\omega \left( \omega -ih\right) }{\left(
h^2+\omega ^2\right) }\right) ^{-1}
$$
and
\begin{widetext}
$$
\rho (\omega ^2)=-\frac 1\pi \text{Im}\left( \underline{\underline{G}}%
(\omega ^2)\right) _{00}=\frac{\omega \gamma h^2\left( h^2+\omega ^2\right)
}{\pi \left( \left( \left( h^2+\omega ^2\right) (h^2+\omega _b^2)-\gamma
h\omega ^2\right) ^2+\omega ^2\gamma ^2h^4\right) }
$$
\end{widetext}
One should notice that

\begin{widetext}
\begin{equation*} \lim _{h\rightarrow \infty }\frac{\omega
\gamma h^2\left( h^2+\omega ^2\right) }{\pi \left( \left( \left(
h^2+\omega ^2\right) (h^2+\omega _b^2)-\gamma h\omega ^2\right)
^2+\omega ^2\gamma ^2h^4\right) }=\frac{\gamma \omega }{\pi \left(
(\omega ^2+\omega _b^2)+\gamma ^2\omega ^2\right) }=\rho (\omega
^2)_{H=\frac 12}
\end{equation*}
\end{widetext}

Thus in this limit the result obtained is identical with the friction $%
\gamma (t)=\gamma 2H\left( 2H-1\right) \left| t\right| ^{2H-2}$ with $%
H=\frac 12$. In other words the Markovian dynamics is recovered.

To calculate $\kappa(t)$ we proceed as in the previous case.

First considering the case $\omega^2=-\Lambda^2$, one gets
\begin{equation*}
\frac {2 \omega ^2}{M \pi} \int\limits_0^\infty \frac{J(\tilde \omega
)d\tilde \omega }{\left( \alpha ^2+\tilde \omega ^2\right) \left( \omega
^2+i\eta -\tilde \omega ^2\right) }=\frac{\gamma h\Lambda }{\left( h+\Lambda
\right) }
\end{equation*}

Thus

$$
\left( \underline{\underline{G}}(\omega ^2)\right) _{00}=\left( -\Lambda
^2+\omega _b^2-\frac{\gamma h\Lambda }{\left( h+\Lambda \right) }\right)
^{-1}
$$

Hence to find out the poles one has to solve the solution of the equation

\begin{equation}
\label{g00lambdakohen}-\Lambda ^2+\omega _b^2-\frac{\gamma h\Lambda }{\left(
h+\Lambda \right) }=0
\end{equation}

To explore the non-Markovian dynamics we choose three set of
parameters corresponding to three different regimes of non-Markovian
dynamics following Kohen and Tannor \cite{TannorJCP1995,
TannorACP2000}. These regimes are called non-adiabatic, caging and
intermediate according to Kohen and Tannor \cite{TannorJCP1995,
TannorACP2000}. The time dependent coefficients in these regimes are
calculated numerically and plotted against time. We observe similar
kind of oscillatory behavior of $\kappa(t)$ in the caging regime as
obtained by them. Similarly in the intermediate regime the plot
looks similar with a nonmonotonic decay (Fig. 2).

\section{Conclusions}

The paper shows an elegant way of combining the traditional system plus
reservoir model \cite{WeissBook, ZwanzigBook} and the method of reactive
flux \cite{ChandlerJSP1986, ChandlerBook} to calculate the time dependent
transmission coefficient, $\kappa(t)$. Analytically we derive a general
formula for $\kappa(t)$ Eq. (\ref{kappat}). For the Markovian case with no
memory ($H=\frac 12$) for the friction $\gamma (t)=\gamma 2H\left(
2H-1\right) \left| t\right| ^{2H-2}$ with $H=\frac 12$ which is same as with
$h=\infty$ for the friction $\gamma (t)=\gamma h\exp \left( -ht\right) $, $%
\kappa(t)$ is calculated analytically. Whereas in the case with memory ($%
H=\frac 34$) and with any finite value of $h$ the calculations were
done numerically. In all the cases a plot of $\kappa(t)$ vs $t$
starts from $1$ at $t=0$ and reaches a plateau in the long time
limit. As expected our formula for the time dependent rate constant,
$k(t)$ becomes equal to the transition state rate constant when one
takes the limit $t \rightarrow0$ \cite {TannorJCP1995,
TannorACP2000}. Similarly Kramers rate constant is obtained by
taking $t \rightarrow\infty$ \cite {TannorJCP1995, TannorACP2000}.
In future we would like to extend our formulation to quantum domain.

\section{Acknowledgements}

The author is grateful to Prof. K. L. Sebastian for his useful
comments and encouragements and thanks Prof. B. J. Cherayil for his
comments on the manuscript. The author also acknowledges Council of
Scientific and Industrial Research (CSIR), India for financial
support.
\\
\\
\section{ Appendix : Transition State Rate and Kramers Rate from $k(t)$}

First we take the limit $t \rightarrow0$. Then from Eq. (\ref{ktsimple}) one
writes

\begin{widetext}
$$
\begin{array}{c}
k(0)=\lim _{t\to 0}k(0+t) =\frac 1{2\pi } \frac{\omega _0}{\omega
_b}\lim _{t\rightarrow 0}\frac{-\dot c(0+t)}{\left(
c(0+t)^2-c(0)^2\right) ^{\frac 12}}\exp (-\beta E_b) =\\ \\
\frac1{2\pi }
\frac{\omega _0}{\omega _b}\lim _{t\rightarrow 0}\frac{-\dot c(0)-\ddot c(0)t}{\left( \left( c(0)+t\dot c(0)+\frac{t^2}2\ddot c(0)\right)
^2-c(0)^2\right) ^{\frac 12}}\exp (-\beta E_b) \\  \\
=\frac 1{2\pi } \frac{\omega _0}{\omega _b}\lim _{t\rightarrow
0}\frac{-\ddot c(0)t}{\left( c(0)^2-c(0)^2+c(0)\ddot c(0)t^2\right)
^{\frac 12}}\exp (-\beta E_b) =\frac 1{2\pi } \frac{\omega
_0}{\omega _b}\frac{-\ddot c(0)}{\sqrt{c(0)\ddot c(0)}}\exp
(-\beta E_b)  =\frac 1{2\pi }\frac{\omega _0}{\omega _b}\frac 1{\sqrt{-c(0)}}\exp (-\beta E_b)
\end{array}$$
\end{widetext}

Where we have used the facts, $\dot c(0)=0$ and $\ddot c(0)=-1$ . This is
because $\dot c(t)=-2\int\limits_0^\infty d\omega \rho (\omega ^2)\sin
(\omega t)$ making $\dot c(0)=0$. Similarly $\ddot
c(t)=-2\int\limits_0^\infty d\omega \rho (\omega ^2)\sin (\omega t)\omega
=-\sum\limits_{j=0}^NU_{j0}^2=-1$. Now one can also show, $-c(0)=\frac
1{\omega _b^2}$ as follows

\begin{widetext}
\begin{equation*}
\begin{array}{c}
c(0)=\sum\limits_{j=0}^N \frac{U_{j0}^2}{\lambda _j^2}
=\sum\limits_{j =0}^NU_{j 0}\lambda _j ^{-2}U_{0 j }=\sum\limits_{j
=0}^N\left\langle 0\mid j \right\rangle \lambda
_j ^{-2}\left\langle j \mid 0\right\rangle \\
\\
=-\lim _{\omega ^2\rightarrow 0}\sum\limits_{j =0}^N\left\langle
0\mid j \right\rangle \left\langle j \right| \frac 1{\left( \omega
^2-D\right) }\left| j \right\rangle \left\langle j \mid
0\right\rangle =-\lim _{\omega ^2\rightarrow 0}\sum\limits_{j
=0}^N\left\langle 0\right| \frac 1{\left( \omega ^2-D\right) }\left|
0\right\rangle =-\lim _{\omega ^2\rightarrow 0}\left(
\underline{\underline{G}}(\omega ^2)\right) _{00}%
\end{array}
\end{equation*}
\end{widetext}

From Eq. (\ref{g00j}) one can see that $\lim _{\omega ^2\rightarrow 0}\left(
\underline{\underline{G}}(\omega ^2)\right) _{00}=\frac 1{\omega _b^2}=-c(0)$%
. Then $k(0)$ is nothing but the transition state rate constant, $k_{TST}$.

\begin{equation}
\label{ktst}k(0)=\frac{\omega _0}{2\pi }\exp (-\beta E_b)=k_{TST}
\end{equation}

Similarly taking $t \rightarrow\infty$ one gets the Kramers rate constant as
follows
\begin{widetext}
\begin{equation*}
\begin{array}{c}
k(\infty )=\frac 1{2\pi } \frac{\omega _0}{\omega _b}\left( \frac{\frac{U_{00}^2}\Lambda \sinh (\Lambda t)}{\sqrt{\left( \frac{U_{00}^2}{\Lambda ^2}\cosh (\Lambda t)\right) ^2}}\right) _{t\rightarrow \infty }\exp (-\beta
E_b) =\frac 1{2\pi }\frac{\omega _0}{\omega _b}\left( \Lambda \tanh (\Lambda
t)\right) _{_{t\rightarrow \infty }}\exp (-\beta E_b)%
\end{array}%
\end{equation*}
\end{widetext}
Then
\begin{equation*}
k(\infty)=\Lambda \frac 1{2\pi }\frac{\omega _0}{\omega _b}\exp (-\beta E_b)
\end{equation*}

Now if one replaces $\Lambda$ by $\Lambda _{H=\frac 12}=-\frac \gamma 2+%
\sqrt{\left( \frac \gamma 2\right) ^2+\omega _b^2}$ ($H=\frac 12$ means
white noise) one gets

\begin{widetext}
$k(\infty)=\left(-\frac \gamma 2+\sqrt{\left( \frac \gamma 2\right)
^2+\omega _b^2}\right)\frac 1{2\pi }\frac{\omega _0}{\omega _b}\exp
(-\beta E_b)$ \end{widetext}

Which is nothing but the usual Kramers rate. Also in the overdamped limit
i.e. when $\gamma>>\omega_b^2$ one gets

\begin{widetext}
\begin{equation*}
k(\infty )=\left( -\frac \gamma 2+\frac \gamma 2\left( 1+\frac
124\left( \frac{\omega _b^2}{\gamma ^2}\right) \right) \right)
\frac{\omega _0}{2\pi \omega _b}\exp (-\beta E_b)=\frac{\omega
_0\omega _b}{2\pi \gamma }\exp (-\beta E_b)
\end{equation*}
\end{widetext}

which is Kramers rate expression in overdamped regime.

\end{document}